\documentclass[article,14pt]{article}
\usepackage{graphicx}
\usepackage{amssymb}
\usepackage{amscd}
\usepackage{amsmath}
\usepackage[english]{babel}
\usepackage[T2A]{fontenc}
\usepackage{EuScript}
\usepackage{eufrak}
\usepackage{float}

\newcommand{\bee}{\begin{eqnarray}}
\newcommand{\eee}{\end{eqnarray}}

\def\beq{\begin{equation}}
\def\eeq{\end{equation}}
\def\ba{\beq\begin{array}{c}}
\def\ea{\end{array}\eeq}

\newcommand{\labcd}[2]{\hbox to\textwidth{#1\dotfill #2}}

\begin{document}


\centerline{\Large{{\bf  Infrared Sensitivity of Unstable Vacua }}} \vspace{%
1.5cm} \centerline{Dmitry Krotov, Alexander M.
Polyakov}\vspace{1.5cm}

\begin{center}
Joseph Henry Laboratories, Princeton University, Princeton, NJ
08544, USA
\end{center}

\vspace{1.5cm}

\centerline{ABSTRACT} \bigskip

We discover that some unstable vacua have long memory. By that we
mean that even in the theories containing only massive particles,
there are correllators and expectation values which grow with time.
We examine the cases of instabilities caused by the constant
electric fields, expanding and contracting universes and, most
importantly, the global de Sitter space. In the last case the
interaction leads to a remarkable UV/IR mixing and to a large back
reaction. This gives reasons to believe that the cosmological
constant problem could be resolved by the infrared physics.

\vspace{1cm}

\thispagestyle{empty}

\section{Introduction}

The problem of cosmological constant presents a serious challenge to
modern physics. Recently the following physical mechanism was
proposed to overcome this problem \cite{AM-recent}. Let us imagine
that the bare cosmological constant is present in the lagrangian. As
is well known, it is the cause of gravitational repulsion, resulting
in the accelerated expansion of the universe (the contraction is
also possible, but we will discuss it later). According to the
proposal, this acceleration leads to the explosive particle
production. The gravitational attraction between the particles slows
down the acceleration and thus reduces (asymptotically to zero) the
effective cosmological constant.

There are many puzzles associated with this proposal. Particles in
the curved space are ill defined, does it make sense to ascribe to
them a real physical effect? Even if it does, the Universe is
exponentially expanding, so it may seem that the particles get
diluted; isn't their back reaction negligible? How can massive
particles considered in \cite{AM-recent} lead to large infrared
effects?

In this paper we will try to provide some clarifications, as well as
present some new results. It is helpful to consider various cases of
unstable vacua and to make their comparative studies. The basic
origin of the difficulties lies in the non-equilibrium quantum field
theory and are common to all the cases. Let us demonstrate this with
the following example.

Consider a~nucleus with a charge $Z$ and hit it with a
$\gamma$-quantum of energy $\omega$ which produces a pair~$e^+e^-$.
The electron forms a~bound
state while the positron escapes to infinity. The~threshold singularity in~$%
\omega$ is located at
\begin{equation*}
\omega=2m-|E_B|
\end{equation*}
where $|E_B(Z)|$ is the~binding energy. We see that when the~nucleus
becomes supercharged, $|E_B(Z)|\!=\!2m$, we get a long-ranged
correlations in time since the threshold is now at~$\omega\!=0$.

Below we will find that such "long-memory" is a crucial factor in
the non-equilibrium dynamics. In field theory it leads to
the~"adiabatic catastrophe" \cite{Polyakov} and to the obstruction
to Wick's rotation. In the next section we shall briefly summarize
the situation.

Another puzzle mentioned above is the dilution of particles in the
expanding universe, the size of which grows as $a(t)\sim e^{t}$.
However, in any physical quantity, this effect always cancels with
the exponentially growing number of the comoving modes. The
covariant cut-off for the comoving momentum $k$ is given by $k\leq
k_{\max }\sim M_{planck}a(t)$, and this causes the growth. The naive
reason for the above compensation is that the change $a\rightarrow
\lambda a$ is a coordinate transformation and $a$ dependence can't
be physical. As will be discussed below, there are caveats to this
argument, but on a qualitative level they are unimportant.

What about the infrared effects generated by the massive particles?
They are not related to the interaction of these particles, which is
short-ranged as usual. Their origin lies in the fact that the
original vacuum is unstable with the non-zero decay rate. Therefore
we get perturbative corrections containing secular terms, which
represent the fact that, as the time goes, it is less and less
likely for the vacuum to remain intact.

\section{Lorentzian vs Euclidean calculations}

A crude example of these phenomena is provided by~a~hot plasma
in~a~box. One can calculate various correlations using the~Euclidean
approach by~introducing Matsubara's periodic time. The~resulting
physical quantities will describe a~well-defined physics of plasma
kept at fixed temperature by~the~external sources. However, if such
sources are absent, the~plasma will cool off. In~this case
the~"Euclidean" calculation is inadequate and we have to~use
the~Schwinger-Keldysh approach.

Similar situation exists in~the~$dS$ space. This space is defined
by~the~equation
\begin{equation*}
\vec{n}^2-n_0^2\ =\ 1
\end{equation*}
It is tempting to define the following rules of the game. Make
a~"Wick rotation", $n_0\Rightarrow in_0$, which transforms the~$dS$
space into a sphere. On a sphere massive particles will never
generate any IR divergences, at least perturbatively. Suppose now
that we calculate the correlation functions
$<\!\!\varphi(n_1)...\varphi(n_N)\!\!>$ on a sphere and then
continue them back to~the~$dS$ space. Such a strategy was briefly
considered and rejected in \cite{AM-recent}, but taken as a
fundamental definition of the theory in \cite{Marolf-Morrison}.
Moreover, it was shown in~these two papers that the analytic
continuation from a sphere doesn't lead to any pathologies.

Our claim is that, just as in the above case of plasma, this
"Euclidean" approach describes the~de~Sitter space artificially kept
at~fixed Gibbons-Hawking temperature. This can be achieved
by~surrounding the~$dS$ patch by reflecting walls sending all
emitted radiation back. However without these magic devices
Euclidean approach is inadequate.

A very similar situation occurs in the case of the Schwarzschild
black holes
--- their Euclidean geometry is the well-known cigar and the field theory on
the cigar describes eternal black hole in which the Hawking
radiation is being returned to~keep the equilibrium. Let us notice
also that in the case of the black hole the Euclidean approach uses
$r\!\geq\!2M$ region where (after Wick's rotation) the killing
vectors are positive.

As is well known, its Euclidean metric%
\begin{equation*}
ds^{2}=(1-\frac{2M}{r})dt^{2}+\frac{dr^{2}}{(1-\frac{2M}{r})}+r^{2}d\Omega
\end{equation*}%
is complete and non-singular if $t\sim t+8\pi M.$ It has a geometry
of a cigar. According to Gibbons and Hawking, periodicity in
imaginary time is the indication of Hawking's temperature. An
interacting field theory on this manifold is also well defined. We
can get a set of Green's functions by an analytic continuation to
real time.

However, as well known, these continuation will describe the black
hole on "life support" - its temperature must be kept constant. The
real black hole evaporates and can't be described by Wick's
rotation. The same is true for the dS space. A typical puzzle here
is that while the black hole evaporates into the outer space, the dS
space has nowhere to go. In fact, this puzzle is psychological. The
dS space simply creates an avalanche of particles within itself .
One can also visualize this by immersing a large patch of the dS
space in the Minkowski space-time. In this case the created
particles will populate the ambient space. For the proper
description of the $dS$ space one must use the Schwinger-Keldysh
approach with real time.

 In the case of the~Bunch-Davies vacuum there are some
further puzzles. The wave functional in this case is
of~the~Hartle-Hawking type:
\begin{equation*}
\Psi_0[\varphi(\vec{n})]=\int\limits_{\varphi|_{\partial
M}=fixed}D\varphi\, e^{-S(\varphi)}
\end{equation*}
where we integrate over fields on a hemisphere. The correlators are
given by
\begin{equation*}
G=(\psi_0,\ \varphi(n_1),..., \varphi(n_N)\,\psi_0).
\end{equation*}

The~$dS$ vector $n$ can be parameterized as $n=(\sinh t, (\cosh
t)\vec{n})$, the southern hemisphere is described by
$-i\pi/2\leqslant t\leqslant0$, while the northern one corresponds
to $i\pi/2\geqslant t\geqslant0$. If we consider a perturbation
theory analytically continued to~$dS$ we have formally
\begin{equation*}
G_{(Euclidean)}\ =\ \Big(\psi_0,\ T_C\varphi(n_1)...\varphi(n_N)\ e^{-\frac{%
i\lambda}{4!}\int\varphi^4}\psi_0\Big)
\end{equation*}
where the contour $C$ for the correllators on a sphere is shown at
Fig.1 left, while the contour after analytic continuation to $dS$
space is shown on the right.
\begin{figure}[h]
\center{\includegraphics[width=130mm]{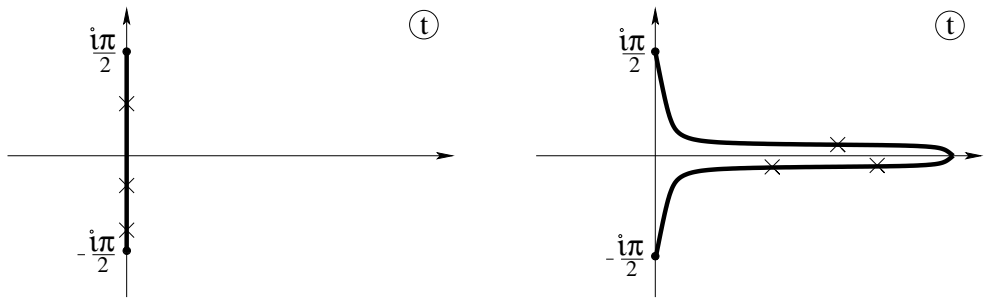}}
\caption{{\protect\footnotesize The contours of integration on a
sphere and after analytic continuation to $dS$ space.}}
\end{figure}
These contours must be used in the~Schwinger-Keldysh diagrams. One
might think that this should give the~same result as the~standard
Schwinger-Keldysh contour for the~Lorentzian approach, Fig.2.
\begin{figure}[h]
\center{\includegraphics[width=50mm]{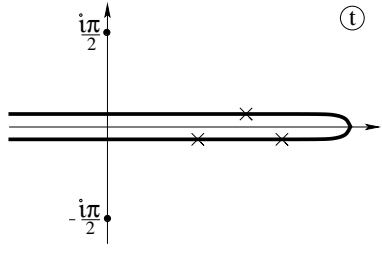}}
\caption{{\protect\footnotesize Standard contour of integration in
Lorentzian approach.}}
\end{figure}

This is not the case. The reason for the discrepancy lies in the
adiabatic catastrophe. We have to introduce the adiabatic switching
of the interaction and let the system develop freely:
\begin{equation*}
G_{(Lorentzian)}(n_1,...)=\langle0|T\varphi(n_1)...\varphi(n_N)e^{-i\int\lambda(%
\varepsilon n_0)\varphi^4dn}|0\rangle
\end{equation*}
where $\varepsilon\rightarrow0$.

As we will show below the IR cut-off $\varepsilon$ \textbf{does not}
disappear and that implies the instability of~the~$dS$ space.
The~reason for~the~breakdown of adiabaticity is that there are
states with negative energies and so the~arbitrary slow external
field can trigger pair production.

\section{Electric fields and new anomaly}

Pair production by electric fields has been discussed in the
hundreds of valuable papers, see references in \cite{Electric
Field}. We return to this problem for two reasons. First, we need to
present it in the form, which can be easily generalized to the
gravitational case. Second, we will find an unusual anomalous vacuum
polarization which may have unexpected applications.

Let us consider massive scalar field in electric field, described by
a time-dependent vector potential $A_1(t)$. We assume that electric
field is
switched on and off adiabatically. This means that it has the form $E\ =\ E(%
\frac{t}{T})$ so that for $|t|<<T$, it remains constant while for $|t|>>T$, $%
E\rightarrow 0$. A good concrete example of such behavior (already
considered in \cite{Nikishov}) is to take
\begin{equation*}
A_1(t)\ =\ ET \tanh\Big(\frac{t}{T}\Big)
\end{equation*}
$E(t)\ =\ \frac{E}{\cosh(\frac{t}{T})^2}$, but explicit shape of the
potential is not important. The Klein-Gordon equation has the form
\begin{equation}  \label{Klein-Gordon-electric-field}
\Big( \partial_t^2\ +\ \big(k-A(t)\big)^2\ +\ k_{\bot}^2\ +\ m^2 \Big) %
\varphi\ =\ 0
\end{equation}
We are interested in the 'in' solution which is defined as the 'Jost
function', i.e. it has the asymptotic behavior
\begin{equation}\label{in-early times}
\varphi_{in}(t,k)\ \rightarrow_{t\rightarrow-\infty}\
\frac{1}{\sqrt{ 2\omega_k^-}} e^{-i\omega_k^- t}
\end{equation}
where $\omega_k^{\pm}\ =\ \sqrt{\big(k-A(\pm \infty)\big)^2\ +\
k_{\bot}^2\
+\ m^2}$. The solution is normalized by the condition that Wronskian $%
W(\varphi, \varphi^\ast)\ =\ 1$.

As we go to the late time $t\rightarrow\infty$, we have
\begin{equation}\label{late time phi-in}
\varphi_{in}(t,k)\rightarrow_{t\rightarrow\infty} \frac{1}{\sqrt{2\omega_k^+}%
}\Big[ \alpha(k)e^{-i\omega_k^+ t}\ +\ \beta(k)e^{i\omega_k^+ t}
\Big]
\end{equation}
where $\alpha$ and $\beta$ are Bogolyubov coefficients also related
to the transmission and reflection amplitudes.

If we start with $\varphi_{in}$ and blindly apply the WKB
approximation, we get
\begin{equation}\label{quasiclassics}
\varphi_{in}(t,k) \ \sim\ \frac{1}{\sqrt{2\omega_k(t)}}
e^{-i\int\limits_0^t \omega_k(t^{\prime })dt^{\prime }}
\end{equation}
for late times, with $\omega_k(t)\ =\ \sqrt{\big(k-A(t)\big)^2\ +\
k_{\bot}^2\ +\ m^2}$. Of course in this way we loose the over
barrier reflection and thus the above formula can't be valid
everywhere. Indeed the WKB requires that the de Broglie wave length
$\lambda=\frac{1}{\omega_k}$ satisfies
\begin{equation*}
\gamma\ =\ \frac{d\lambda}{dt}\ =\ \frac{\big(k-A\big)\dot{A}(t)}{\Big[%
(k-A)^2\ +\ k_{\bot}^2\ +\ m^2\Big]^{\frac{3}{2}}}\ \ll\ 1
\end{equation*}
We see that the approximation is good for the early times when
$|k-A(t)|\ \gg\ m$. In this case
\begin{equation*}
\gamma\ \sim\ \frac{E}{|k-A|^2}\ \sim\
\frac{m^2}{|k-A|^2}\frac{E}{m^2}\ \ll\ 1
\end{equation*}
if we assume $E\ \sim m^2$.

However, around the point where the mode 'reaches the horizon', defined by $k=A(t_k)$, we get $%
\gamma\ \sim\ 1$ and WKB breaks down. As we go to $t\gg t_k$,
$|k-A|$ starts growing again and the WKB is valid again. In this
region it must contain two exponentials:
\begin{equation}\label{in-late times}
\varphi_{in}(t,k)\ \sim \ \frac{1}{\sqrt{2\omega_k(t)}} \Big[ %
\alpha(k)e^{-i\int\limits_0^t \omega_k}\ +\
\beta(k)e^{i\int\limits_0^t \omega_k} \Big] \ \ \ \ \ \ \ \ \ \ \ \
\ t\gg t_k
\end{equation}
As usual, $\alpha$ and $\beta$ can be found by matching
(\ref{in-early times}) and (\ref{in-late times}).

In the domain $|t|\ll T$ the electric field is constant and
$A(t)\sim E t$.
The equation (\ref{Klein-Gordon-electric-field}) now depends on the variable $%
t-\frac{k}{E}$, hence $\varphi_{in} \sim f_{in}(t-\frac{k}{E})$. The
function $f_{in}$, as well known, is the parabolic cylinder function
\begin{equation*}
\varphi_{in}\ \sim\ D_{-\frac{1}{2}\ -\ i\lambda}\Big[ - \sqrt{2 E}
e^{ i\frac{\pi}{4}}(t\ -\ \frac{k}{E}) \Big]\ \ \ \ \ \ \ \ \
t\rightarrow-\infty
\end{equation*}
but we will not need their explicit form. What is important is that
due to the symmetry $k\rightarrow k+\kappa$, $t\rightarrow
t-\frac{\kappa}{E}$ the resulting $\alpha$ and $\beta$ do not depend
on $k$ in a certain range, which we determine in a moment (but do
depend on $k_\bot$ and $m$).

To find this range we notice that the 'horizon crossing' ($k=A(t)$)
occurs
at $t_{k}=\frac{k}{E}$. We can use the constant field approximation only if $%
t_{k}\ll T$. Hence we conclude that $\alpha $ and $\beta $ do not
depend on $k$ only if $A(-\infty )<k<A(\infty )$. Outside this
interval reflection coefficient $\beta$ quickly decreases to zero.

The field $\varphi $ can be expanded in terms of creation and
annihilation operators as
\begin{equation*}
\varphi \ =\ \sum\limits_{k}\Big(a_{k}f_{k}^{in\ast }e^{ikx}\ +\
b_{k}^{\dagger }f_{k}^{in}e^{-ikx}\Big)
\end{equation*}%
and the Green function is equal to
\begin{equation*}
\begin{split}
G(x_{1},t_{1}|x_{2},t_{2})\ =\ _{in}\langle 0|T\varphi
(x_{1},t_{1})\varphi (x_{2},t_{2})^{\ast }|0\rangle _{in}\ =\ \int
f_{k}^{in}(t_{<})f_{k}^{in\ast
}(t_{>})e^{ik(x_{1}-x_{2})}dk\ =\\ =\ e^{iE\frac{t_{1}+t_{2}}{2}%
(x_{1}-x_{2})}g(t_{1}-t_{2},x_{1}-x_{2})
\end{split}
\end{equation*}%
The first factor here is a gauge dependent phase which must cancel
in physical quantities. The remaining part
$g(t_{1}-t_{2},x_{1}-x_{2})$ is invariant under translations and
defines correlation functions of gauge invariant quantities. In
particular we can calculate the induced current which can be used to
estimate the back reaction. The general formula for the current is
\begin{equation*}
\langle J(t)\rangle \ =\ \int \big(k-A(t)\big)|\varphi
_{in}(k,t)|^{2}\ dk
\end{equation*}%
As we will see, the current is dominated by the two semi-classical
domains described above. Before the 'horizon crossing' we have
\begin{equation}\label{current-1}
\langle J(t)\rangle ^{(1)}\ =\ \int\limits_{A(t)<k}dk\ dk_{\bot }\frac{%
(k-A(t))}{2\omega _{k}(t)}\ =\ \int\limits_{0}^{\infty }\frac{dp\ p\
dk_{\bot }}{2\sqrt{p^{2}+k_{\bot }^{2}+m^{2}}}
\end{equation}
where $p=k-A(t)$ is 'physical momentum'. After horizon crossing, we
have to use (\ref{late time phi-in}). Keeping only non-oscillating
terms, which are dominant, we obtain
\begin{equation*}
\langle J\rangle ^{(2)}\ =\ \int\limits_{k<A(t)}dk\ dk_{\bot }\frac{k-A(t)}{%
\sqrt{(k-A)^{2}+k_{\bot }^{2}+m^{2}}}\big(|\alpha (k)|^{2}+|\beta (k)|^{2}%
\big)
\end{equation*}%
Using the general relation $|\alpha (k)|^{2}-|\beta (k)|^{2}=1$ we
get:
\begin{equation*}
\langle J\rangle ^{(2)}\ =\ \int\limits_{-\infty }^{0}\frac{dp\ dk_{\bot }\ p%
}{2\sqrt{p^{2}+k_{\bot }^{2}+m^{2}}}\ +\ 2\int\limits_{-\infty }^{0}\frac{%
dp\ dk_{\bot }\ |\beta (k,k_{\bot })|^{2}\ p}{2\sqrt{p^{2}+k_{\bot
}^{2}+m^{2}}}
\end{equation*}%
The first term in this formula combines with (\ref{current-1}) and gives zero due to $%
p\rightarrow -p$ symmetry. The second term is really interesting.
The key feature of it is that the reflection coefficient $\beta $
depends on the 'comoving' momentum $k$ and not the physical one $p$.
As we saw, this coefficient keeps being constant for $A(-\infty )\ll
k\ll A(\infty )$ and quickly vanishes outside this interval. In
terms of $p$, this means the time-dependent cut-off $A(-\infty )\ll
p+A(t)\ll A(\infty )$. We also have a cut-off on $k_{\bot }$,
$k_{\bot }\ll E$. Hence, the total current is given by
\begin{equation}\label{final answer for current}
\langle J\rangle \ =\ \!\!\!\!\!\!\!\!\!\!\int\limits_{\ A(-\infty
)-A(t)}^{0}\!\!\!\!\!\!\!\! dp\ \frac{p}{|p|} \int dk_{\bot }\
|\beta (k_{\bot },k)|^{2}\ =\ -\Big(A(t)-A(-\infty )\Big)
 |\beta
|^{2}
 E^{\frac{d-1}{2}}\cdot const
\end{equation}%
In the last expression $|\beta|^2\ =\ e^{-\frac{\pi m^2}{E}}$. This
result is physically transparent. It means that, as the time goes
by, more and more $k$ modes cross the horizon $k=A(t)$ and begin to
contribute to the induced current. This fact is important. It shows
that the induced current is proportional to the vector potential and
not the field strength. Together with gauge invariance this implies
highly non-local behavior. Indeed, $A(t)-A(-\infty )\ =\
\int\limits_{-\infty }^{t}dt^{\prime }E(t^{\prime })$. Similar
non-localities are well known - the London equation in
superconductors, the photon mass in the Schwinger Model, the
Chern-Simons terms in the quantum Hall effect. In all these cases
the gauge invariant expressions can't be expressed locally in terms
of the the field strengths.

The stunning feature of the above result is that the non-locality
appears in the massive theory. This is specific for unstable vacua
and can't be seen in the in/out formalism. It also implies the
strong back reaction, since the current is growing with time.
Another interpretation of this result is symmetry breaking. Indeed,
in the constant field we obviously have the time translation
invariance. This invariance is broken in the expression for the
current due to the influence of the past when the field was turning
on. We will return to this phenomenon later, while discussing the
gravitational case.

In the mean field approximation we plug the current back into
Maxwell's
equation similarly to the procedure of \cite{Cooper}.\footnote{%
In the recent paper \cite{Emil-Burda} the in/in current has been
calculated. The result is different from ours and we disagree with
the method used in this paper. Another expression for the in/in
current, which is consistent with our result, can be found in
\cite{Gitman}. Let us also notice the paper \cite{Starobinsky} on
R-N black holes which used the method very close to ours.}
Considering for simplicity 1+1 dimensions we get
\begin{equation*}
\ddot{A}=J=-2|\beta (\dot{A})|^{2}A(t)
\end{equation*}%
At large time the solution decays as $E=\dot{A}\ \sim \
\frac{1}{\log t}$ indicating the total screening of the electric
field. This may be related to the old result by Gribov
\cite{Gribov}. Another analogy is the Landauer conductance in the
mesoscopic system, which is also expressed in terms of the
transmission and reflection coefficients.

Another interesting point is that for massless fermions the anomaly
equation reads as
\begin{equation*}
\partial _{0}J_{1}\ =\ \frac{1}{\pi }E
\end{equation*}%
since axial and vector currents are related by epsilon symbol in two
dimensions. This gives linear growth of current with time and is
consistent with our result since in massless case $\beta
\ \sim 1$. The back-reaction can be estimated from the Maxwell equation $%
\dot{E}+J_{1}\ =\ 0$ and is clearly significant. In the case of
fermions (the Schwinger Model) we know that in the vacuum no
electric field remains and electric charges are completely screened.

We can also use the in/out Green function
\begin{equation*}
G^{in/out}\ =\ \frac{1}{\alpha }\varphi _{k}^{in}(t_{<})\varphi
_{k}^{out\ast }(t_{>})
\end{equation*}%
The sign of vacuum instability here is $ImG(t|t)\neq 0$. Let us
notice that the matrix element $\langle out|J_{1}|in\rangle \ =\ 0$
simply because the in/out Green function is Lorentz invariant
(module a phase factor). The Euclidean version of this phenomenon
(with the replacement of the electric field by the magnetic one) is
the absence of bulk currents in the quantum Hall effect.

It is also instructive to change the gauge. If we take $A_{0}\ =\
Ez$ we get Klein-Gordon equation
\begin{equation*}
(\partial _{z}^{2}\ +\ (\omega -Ez)^{2}\ -\ m^{2})\varphi \ =\ 0
\end{equation*}%
As in the time-dependent gauge, we have a Schrodinger equation for
inverted oscillator, but this time the effect of pair creation comes
from the underbarrier penetration rather than from the overbarrier
reflection. The two are related by the analytic continuation. In
this gauge the energy $\omega \ =\ Ez\ +\ \sqrt{p^{2}+m^{2}}$ is
conserved but non-positive which allows particle production. This
gauge has de Sitter counterpart. The tunneling above is analogous to
the tunneling in the Painleve coordinates \cite{Volovick}.

\section{Expanding Universe (free fields)}
When we look at the de Sitter space, we find that there are striking
similarities with the electric case. Let us consider what happens
when the curvature of $dS$ space is adiabatically switched on. In
this setting we have two quite different problems - expanding and
contracting universes. The arrow of time is set up by defining the
infinite past as a Minkowski space in which our field is in the
ground state and solutions to the wave equation are chosen to be the
Jost functions. Let us begin with the expanding Universe.
Analogously to the electric case we will assume that the FRW metric
\begin{equation*}
ds^{2}\ =\ a(t)^{2}d\vec{x}^{\ \!2}\ -\ dt^{2}
\end{equation*}%
is such that $\frac{\dot{a}}{a}\ =\ H(\frac{t}{T})$, time $T$ is supposed to be large, and $H(0)=1$, while $%
H(\pm \infty )=0$. A representative example of such a metric is
\begin{equation*}
a(t)\ =\ e^{T\tanh \frac{t}{T}}
\end{equation*}%
$H(t)\ =\ \frac{1}{\cosh (\frac{t}{T})^{2}}$. It is convenient to
rescale the standard scalar field $\varphi $ by defining $\varphi
=a^{-\frac{d}{2}}\phi $. The Klein-Gordon equation takes the form
\begin{equation*}
\ddot{\phi}_{in}\ +\ \Big(m^{2}\ -\ r(t)\ +\
\frac{k^{2}}{a(t)^{2}}\Big)\phi _{in}\ =\ 0
\end{equation*}%
with $r(t)=\frac{d(d-2)}{4}(\frac{\dot{a}}{a})^{2}+\frac{d}{2}\frac{
\ddot{a}}{a}.$ As before, the 'in' solution is defined by
\begin{equation}\label{in-grav}
\phi _{in}\ =\ \frac{1}{\sqrt{2\omega _{k}^{-}}}e^{-i\omega
_{k}^{-}t}
\end{equation}
as \ $t\rightarrow - \infty $ with $\omega _{k}^{-}\ =\ \Big(m^{2}+\frac{k^{2}%
}{a(-\infty )^{2}}\Big)$. Its quasiclassical expression is given by
the formula (\ref{quasiclassics}) where $\omega
_{k}(t)=\sqrt{m^{2}-r(t)+\frac{k^{2}}{a(t)^{2}}}$. This WKB
expression is applicable if
\begin{equation*}
\gamma \ =\ \dot{\lambda}\ =\ \frac{d}{dt}\Big(\frac{1}{\omega
_{k}}\Big)\
\sim \ \frac{1}{\big(m^{2}-r+\frac{k^{2}}{a^{2}}\big)^{\frac{3}{2}}}\frac{%
k^{2}}{a^{2}}\frac{\dot{a}}{a}\ \ll \ 1
\end{equation*}%
If we assume that $H=\frac{\dot{a}}{a}\ \sim \ m$ and $\dot{H}$ is
small, we see that WKB breaks down when the given mode crosses the
horizon, $k\ \sim \ ma(t)$. Before that we had $k\gg ma(t)$ and
$\dot{\lambda}\ll 1$. Long after that we reach the semi-classical
regime again, but with two exponentials as in (\ref{late time
phi-in}). Let us consider the time evolution of the quantity
$\langle in|\varphi (t)^{2}|in\rangle $. We have:
\begin{equation*}
\langle in|\varphi (t)^{2}|in\rangle \ =\ \int d^{d}k\ |\varphi
_{in}(t,k)|^{2}
\end{equation*}%
splitting the integral as before into the regions $|k|\gg ma(t)$ and
$|k|\ll ma(t)$ we get
\begin{equation}\label{phi^2 expanding universe}
\begin{split}
\langle in|\varphi (t)^{2}|in\rangle \ & =\
a(t)^{-d}\int\limits_{|k|\gg
ma(t)}\frac{d^{d}k}{2\omega _{k}(t)}\ +\ a(t)^{-d}\int\limits_{|k|\ll ma(t)}%
\frac{d^{d}k}{2\omega _{k}(t)}\Big[|\alpha (k)|^{2}+|\beta
(k)|^{2}\Big]\ =
\\
& =\ a^{-d}\bigg(\int \frac{d^{d}k}{2\omega _{k}(t)}\ +\
2\int\limits_{|k|\ll ma(t)}\frac{d^{d}k}{2\omega _{k}(t)}|\beta (k)|^{2}%
\bigg)
\end{split}%
\end{equation}%
The reflection amplitude $\beta (k)$ is $k$-independent in a certain
interval, just as it was in the electric case. The reason is that de
Sitter wave equation is invariant under $k\rightarrow \lambda k$ and
$t\rightarrow t+\log \lambda $ (which is one of the $dS$
isometries). However this amplitude quickly vanishes when $k$ is
such that the horizon crossing happens outside
the de Sitter stage. Namely, if $t_{k}$ is determined from the equation $%
k=ma(t_{k})$, the constant reflection occurs for $|t_{k}|\ll T$. If
we
introduce the cut-offs defined by $\frac{k_{min}}{a(-\infty )}=k_{min}e^{T}=m$ and $%
\frac{k_{max}}{a(+\infty)}=k_{max}e^{-T}=m$, we have reflection only
if $k_{min}\ll k\ll k_{max}$. We see that the contribution of the
second term in (\ref{phi^2 expanding universe}), which represents
the created particles, is small in the expanding case. Due to the
infrared convergence of the integral we obtain
\begin{equation}\label{phi^2 final result}
\langle \varphi (t)^{2}\rangle ^{(2)}\ \sim \ |\beta |^{2}m^{d-1}
\end{equation}%
This formula has a clear physical interpretation. By the moment $t$
we excite the modes with $|k|<ma(t)$ and the average excitation number is $%
\overline{n}\ \sim |\beta |^{2}$. The created particles are
non-relativistic due to the upper boundary on $k$. Let us stress
that there is no dilution of the created particles in the sense that
their physical (not comoving) density remains constant in time,
however their main contribution is just a renormalization of the
cosmological constant which is unobservable.

 The key difference from
the electric case is the absence of the dynamical symmetry breaking,
which we define as a long-term memory. By this we mean the
following. As we already noticed, the current in the electric case
depends on the time passed from the first appearance of the field.
This effect is a dynamical counterpart of the usual spontaneous
symmetry breaking. In the latter case, the magnetic field at the
boundary induces magnetic moment in the bulk, if we talk about Ising
model for example. In our case the role of the boundary is played by
the infinite past. The expression (\ref{phi^2 final result}) does
not depend on time. Hence, there is no dynamical breaking of de
Sitter symmetry in this case. Life becomes more interesting if we
switch on interaction or consider contracting universe.

 We could calculate things in the regime of the constant curvature and get the right results.
In this case
\begin{equation*}
\varphi _{in}\ \sim \ \tau ^{\frac{d}{2}}H_{i\mu }^{(1)}(k\tau )
\end{equation*}%
with $\tau =e^{-t}$ and
\begin{equation*}
\langle in|\varphi (t)^{2}|in\rangle \ \sim \ \tau ^{d}\int d^{d}k\
|H_{i\mu }^{(1)}(k\tau )|^{2}\ =\ \int d^{d}p\ |H_{i\mu
}^{(1)}(p)|^{2}\ =\ const
\end{equation*}%
The UV divergence in this integral is the same as in the flat space
and the time independence in this formula is just the result of the
de Sitter symmetry. The back reaction is thus small and
uninteresting. Really non-trivial things begin to happen when we
either include interactions or consider contracting universe. We
start with the latter.

\section{Contracting Universe (free fields)}
Let us repeat the above calculations in the case of contracting Universe%
\footnote{%
We considerd the case of contracting universe following the advice
of V. Mukhanov.}. At the first glance it may seem that, since de
Sitter space is time-symmetric, expansion and contraction can't lead
to different results. However, as was stated above there is an arrow
of time in our problem. We defined the past by the condition that
our field is in the Minkowsky vacuum state. Generally speaking, in
the future we should expect complicated excited state. In this
setting contraction is very different from expansion. We can once
again take
\begin{equation*}
a(t)\ =\ e^{-T\tanh \frac{t}{T}}
\end{equation*}%
The modes with $k>ma(-\infty )=me^{T}$ will always stay in the WKB
regime,
since $a(t)$ will be decreasing. On the other hand, the modes with $%
ma(\infty )\ll k\ll ma(-\infty )$ will cross the horizon at some time, $%
k\approx ma(t_{k})$. If we once again define the 'in' modes,
$\varphi _{in}(k,t)$ by the condition (\ref{in-grav}), we find that
for $k\ll ma(t)$ the horizon crossing (WKB breaking) has not
occurred yet (remember that $a(t)$ is decreasing) and hence we have
a single exponential (\ref{in-grav}).

For $ma(t)\ll k\ll ma(-\infty )$ the horizon crossing is already in
the past
and we have two exponentials with the coefficients $\alpha $ and $\beta $, $%
|\alpha (k)|^{2}\ -\ |\beta (k)|^{2}\ =\ 1$. For $k\gg ma(-\infty
)$, the horizon crossing has never occurred and $\beta \rightarrow
0$. As in the previous section we get
\begin{equation}
\begin{split}
\langle in|\varphi (t)^{2}|in\rangle \ & =\
a(t)^{-d}\!\!\!\!\!\!\!\!\!\!\!\!\!\!\!\!\int\limits_{|k|\ll ma(t),
\ |k|\gg
ma(-\infty)}\!\!\!\!\!\!\!\!\!\!\!\!\!\!\!\!\frac{d^{d}k}{2\omega
_{k}(t)}\ \ +\ \
a(t)^{-d}\!\!\!\!\!\!\!\!\!\!\!\!\!\!\!\!\int\limits_{ma(t)\ll|k|\ll
ma(-\infty )} \!\!\!\!\!\!\!\!\!\!\!\!\!\!\!\!\frac{d^{d}k}{2\omega
_{k}(t)}\Big[|\alpha (k)|^{2}+|\beta
(k)|^{2}\Big]\ = \\
& =\ a^{-d}\bigg(\int \frac{d^{d}k}{2\omega _{k}(t)}\ \ +\ \
2\!\!\!\!\!\!\!\!\!\!\!\!\!\!\!\!\int\limits_{ma(t)\ll|k|\ll ma(-\infty )}\!\!\!\!\!\!\!\!\!\!\!\!\!\!\!\!\frac{d^{d}k}{2\omega _{k}(t)}|\beta (k)|^{2}%
\bigg)
\end{split}%
\end{equation}%
Collecting different terms we get
\begin{equation}
\begin{split}
\langle \varphi (t)^{2}\rangle \ & =\ a^{-d}\int\limits_{|k|\ll \Lambda a(t)}%
\frac{d^{d}k}{2\omega _{k}}\ \ +\ \ 2|\beta
|^{2}a^{-d}\!\!\!\!\!\!\!\!\!\int\limits_{ma(t)<|k|<ma(-\infty )}\!\!\!\!\!\!\!\!\frac{d^{d}k}{2\omega _{k}(t)}%
\ \approx \\
& \approx \ const\cdot \Lambda ^{d-1}\ +\ |\beta |^{2}\Big(\frac{a(-\infty )%
}{a(t)}\Big)^{d-1}m^{d-1}
\end{split}%
\end{equation}%
The first term in this formula is just the same UV divergent term as
in the Minkowsky space. The heart of the matter is the second term
which displays the symmetry breaking through the long-term memory
(dependence on $a(-\infty )$). However, the memory can't be too
long, since we have a standard UV
cut-off at the Planck mass. Because of it, the above formulae are valid if $%
p=k/a(t)<M_{pl_{{}}}$ and therefore $a(-\infty )/a(t)<M_{pl}/m.$

Let us sum up the above discussion. In the expanding universe the
contribution from the created particles comes from the region $m
a(-\infty)\ll k \ll m a(t)$. No long term memory is present and
time-dependent back reaction is small, of the order of $\Big(\frac{a(-\infty)%
}{a(t)}\Big)^{d-1}$. Created particles are non-relativistic due to
the red shift.

In the case of contracting universe particles come from
$ma(t)<|k|<\min \big(ma(-\infty ),\ M_{pl}a(t)\big)$. They are
ultra-relativistic and their contribution is of the order
$\Big(\frac{a(-\infty )}{a(t)}\Big)^{d-1}\ \rightarrow \ \infty $.
All these conclusions are correct only for non-interacting
particles.

It is also possible to calculate the energy-momentum tensor. We have
\begin{equation*}
T_{00}\ =\ \int d^{d}k\Big((\partial _{0}\varphi )^{2}\ +\ \frac{1}{a(t)^{2}}%
(\partial _{i}\varphi )^{2}\ +\ m^{2}\varphi ^{2}\Big)
\end{equation*}%
In the contracting case the order of magnitude of this quantity is
defined by the integral:
\begin{equation*}
T_{00}\ \sim \ a^{-d}\int \frac{d^{d}k}{2\omega
_{k}}\frac{k^{2}}{a^{2}}\ |\beta|^2\
\sim \ a^{-d-1}\!\!\!\!\!\!\!\!\!\!\!\int\limits_{ma<k<ma(-\infty )}\!\!\!\!\!d^{d}k\ |k|\ |\beta|^2 \ \sim \ m^{d+1}\Big(%
\frac{a(-\infty )}{a(t)}\Big)^{d+1} |\beta|^2
\end{equation*}%
This corresponds to the ultra-relativistic particles with the
equation of state $p\ =\ \frac{1}{d}\varepsilon $. In the expanding
case the contribution to $T_{00}$ comes from a small number of
created non-relativistic particles. In both cases there are no
reasons to believe that created particles are in thermal
equilibrium. Let us also stress that the above formula represents a
non-local contribution to $T_{00}$ similar to (\ref{final answer for
current}). In contrast with this formula, the local contributions
should depend on the quantities taken at the time $t$ only.

\section{Secular interactions and the leading logarithms, Poincare patch}
 In this section we discuss a very peculiar property of the de
Sitter space. Namely, it turns out that the interactions of the
massive particles generate infrared corrections. We start with the
second order of perturbation theory in the case of $\lambda \varphi
^{3}$  interactions (which we choose to simplify notations; the
phenomenon we are after is general and has nothing to do with the
naive lack of the ground state of the above interaction).  We first
calculate the correction to the Green's function $G(\vec{q},\tau
)=\langle in|\varphi (\vec{q},\tau )\varphi (-\vec{q},\tau
)|in\rangle $ where $\vec{q}$ is a comoving momentum in the Poincare
patch and $\tau$ is a conformal time. Our goal is to show that if
the
physical momentum  $p=$ $q \tau \ll \mu,$ there are corrections of the order ($%
\lambda ^{2}\log \frac{\mu}{p})^{n}$ where $\mu$ is the particle
mass; notice also that these logarithms are the powers of the
physical time $t=-\log \tau $.

We are interested in the loop corrections to the one-point function
$\langle \varphi(t)^2\rangle$. The magnitude of this quantity
determines the strength of the backreaction. To find it we have to
use the Schwinger- Keldysh perturbation theory. These methods are
well known and we will add a few explanations to fix notations. Let
us suppress first the momentum dependence and expand $\varphi
=f^{\ast }a+fa^{+}$ , where $f(t)$ are the "in" modes and $a$ is an
annihilation operator. The relevant one-loop diagram is shown at
Fig.3. It's contribution to $G(\vec{q},\tau )=\langle in|\ \varphi
(\vec{q},\tau )\ \varphi (-\vec{q},\tau )\ |in\rangle$ is given by
\begin{equation}\label{one-loop contribution}
\begin{split}
G(\vec{q},\tau)\ =\ -\lambda^2\ f_q^\ast(t)^2\
\int\limits_{-\infty}^t dt_1 dt_2\  f_q(t_1)f_q(t_2) \int
\frac{d^dk}{(2\pi)^d}\ \ f_k(t_<) f_k^\ast(t_>) f_{k+q}(t_<) f_{k+q}^\ast(t_>)\ -\ c.c.\ +\\
+\ 2\cdot\lambda^2\ |f_q(t)|^2\ \int\limits_{-\infty}^t dt_1 dt_2\
f_q(t_1) f_q^\ast(t_2)\ \int \frac{d^dk}{(2\pi)^d}\ f_k(t_1)
f_k^\ast(t_2) f_{k+q}(t_1) f_{k+q}^\ast(t_2)
\end{split}
\end{equation}
In the first line we have the contribution of the $(+/+)$ and
$(-/-)$ diagrams (the signs refer to the points $t_{1,2}$ of the
physical time, or $\tau_{1,2}$ of conformal time at the diagram in
Fig.3), while in the second line we have $(+/-)$ and $(-/+)$
diagrams.
\begin{figure}[h]
\center{\includegraphics[width=60mm]{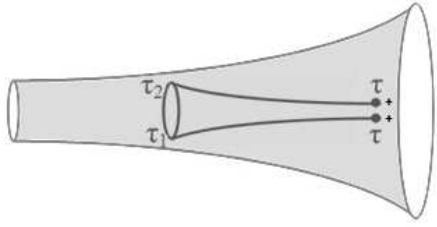}} \caption{{\footnotesize
One-loop diagram responsible for infrared logarithms in Poincare
patch.}}
\end{figure}

\noindent We choose the "in" wave function to be a Hankel function
\begin{equation*}
f_{k}(t)=\tau ^{\frac{d}{2}}h(k\tau )=const\tau ^{d/2}H_{i\mu
}^{(1)}(k\tau )
\end{equation*}%
where the normalization is fixed by the condition $h(x)\rightarrow
(2x)^{-\frac{1}{2}}e^{ix}$ as $x\rightarrow \infty$. With this
normalization, the
asymptotic behavior at $x\rightarrow 0$ is given by%
\begin{equation}\label{small x asymptotics}
h(x)\rightarrow A(\mu )x^{i\mu }+A(-\mu )x^{-i\mu }
\end{equation}
where $A$-s are some concrete functions which we discuss later.

As we will show in a moment, there are infrared logarithmic
corrections to $G(\vec{q},\tau )=\tau ^{d}g(q\tau )$ when
$q\tau\ll\mu$. In this regime we can use asymptotic expressions
(\ref{small x asymptotics}) to get, $\tilde{\lambda}\ =\ \lambda^2
\log\big(\frac{\mu}{q\tau}\big)$
$$
g(x)\ =\ A(\mu)A^\ast(-\mu)\ \Gamma(\tilde{\lambda},\mu)\ x^{2i\mu}\
+\ A(-\mu)A^\ast(\mu)\ \Gamma^\ast(\tilde{\lambda},\mu)\ x^{-2i\mu}\
+\ \big(|A(\mu)|^2+|A(-\mu)|^2\big)\ C(\tilde{\lambda},\mu)
$$
when interaction is off ($\tilde{\lambda}=0$), coefficients
$\Gamma^{(0)}\ =\ C^{(0)}\ =\ 1$. Our goal is to find these
quantities at non-zero $\lambda$. We start with the interference
term $C$.

In order to obtain the contribution to $g(q\tau )$ we have to
integrate the diagrams of Fig.3 over the momentum $k$ and the time
variables $t_{1}$ and $t_{2}$ . The logarithmic contribution comes
from the domain $\tau _{1,2}\sim \mu/k$ and $\mu/\tau \gg k\gg q$.
In this domain we get
the contribution from the first term in (\ref{one-loop contribution}) in the form%
\begin{equation*}
g(q\tau)_{I}=-2\lambda^2\ h^{\ast }(q\tau )^{2}\int d^{d}k\int_{\tau
}^{\infty }d\tau _{1}\int_{\tau _{1}}^{\infty }d\tau _{2}(\tau
_{1}\tau _{2})^{d/2-1}h(q\tau _{1})h(q\tau _{2})h^{\ast }(k\tau
_{1})^{2}h(k\tau _{2})^{2}-c.c.
\end{equation*}%
Taking the limit $q\rightarrow 0$ and interchanging 1 and 2 in the
complex conjugate term  gives
\begin{equation*}
g(q\tau)_{I}=\int_{q}^{\frac{\mu}{\tau} }\frac{d^{d}k}{k^{d}}%
C_{I}(\mu )=C_{I} \log (\frac{\mu}{q\tau })
\end{equation*}%
Here the coefficient is given by%
\begin{eqnarray*}
C_{I} &=&-4\lambda^2 |A(\mu )A(-\mu )|^{2}(|g(\mu )|^{2}+|g(-\mu )|^{2}) \\
g(\mu ) &=&\int_{0}^{\infty }dx\ x^{d/2-1+i\mu}\ h^{2}(x)
\end{eqnarray*}%
The second term is treated analogously. It has the form%
\begin{equation}
g(q\tau)_{II}\ =\ 2\lambda^2 h^{\ast }(q\tau )h(q\tau )\int
d^{d}k\int_{\tau }^{\infty }d\tau _{1}d\tau _{2}(\tau _{1}\tau
_{2})^{d/2-1}h^{\ast }(q\tau _{1})h(q\tau _{2})h^{\ast }(k\tau
_{1})^{2}h(k\tau _{2})^2
\end{equation}
Integration gives another logarithm. Summing these contributions
finally gives for the interference term%
\begin{equation*}
\langle \varphi _{q}^{2}\rangle\ =\ g(q\tau)_I+g(q\tau)_{II}\ =\
2\cdot\Big(B(\mu )-B(-\mu )\Big)\cdot\Big(B(\mu )|g(\mu
)|^{2}-B(-\mu )|g(-\mu )|^{2}\Big)\cdot \lambda^2
\log\Big(\frac{\mu}{q\tau }\Big)
\end{equation*}%
where
\begin{equation*}
B(\mu )\ =\ |A(\mu )|^{2}\ =\ \frac{1}{4\mu}\ e^{\pi\mu}
\frac{1}{\sinh(\pi\mu)}
\end{equation*}
The first multiple here is a Wronskian of the eigenmodes. The second
one turns out to be equal to zero. To see this, note that the
functions $h(x)$ satisfy $ h(x)^\ast\ =\ e^{i\frac{\pi}{2}}\
h(e^{i\pi} x) $ which implies the following relation for $g(\mu)$:
$$
|g(\mu)|^2\ =\ e^{-2\pi\mu} |g(-\mu)|^2
$$
The physical meaning of this equality is detailed balance relation
with Gibbons-Hawking temperature for de Sitter space. Combining this
with the similar property for $A(\mu)$, we conclude that the
one-loop contribution to the coefficient in front of the logarithmic
divergence in the interference term is equal to zero
$C^{(1)}(\tilde{\lambda},\mu)=0$.

The next step is to calculate $\Gamma$. Imaginary part of this
quantity determines the renormalization of mass $\mu$, which we are
not interested in at the moment. The real part is responsible for
the imaginary contribution to $\mu$, which is related to the decay
rate of the particle. Using similar tricks\footnote{It is convenient
to rescale $k$ from the integrals over $\tau_{1,2}$ and note that
$$
Y\ =\ \int\limits_{0}^{\infty} dx \int\limits_x^\infty dy\
(xy)^{\frac{d}{2}-1} \Big(
\big(\frac{x}{y}\big)^{i\mu}+\big(\frac{x}{y}\big)^{-i\mu}  \Big)
h(y)^2 h^\ast(x)^2\ =\ \frac{1}{2}\big(|g(\mu)|^2+|g(-\mu)|^2\big)\
+\ i A
$$ where $A$ is some real number, contributing to renormalization of $\mu$ only.} to those used above we find
$$
Re\Big(\Gamma^{(1)}\Big)\ =\ \lambda^2 \Big(B(\mu)-B(-\mu)\Big)
\Big(|g(\mu)|^2-|g(-\mu)|^2\Big)\log\Big(\frac{\mu}{q\tau}\Big)
$$
This quantity is non-zero and negative.

The above calculation refers to the IR properties of the two-point
function. In the case of Poincare patch there is no IR contribution
to the one-point quantities, as can be seen from the conformal
diagram at Fig.4. The Poincare patch is shown here by the gray area.
Interactions contributing to the one-point function must be located
inside the past light cone due to causality. Therefore we have to
consider only the intersection of the light-cone with the gray area
defining Poincare patch.
\begin{figure}[h]
\center{\includegraphics[width=60mm]{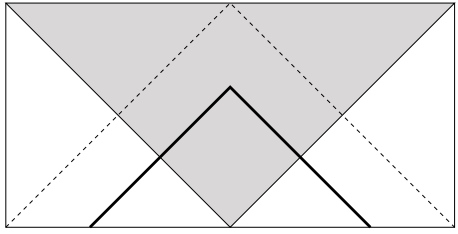}} \caption{{\footnotesize
Conformal diagram. Poincare patch is shown by the gray area. Solid
black line represents the past light cone of the observer. The
intersection of this cone with Poincare patch touches past infinity
only at one point.}}
\end{figure} Thus infrared effects in Poincare patch can not have
dramatic consequences because the past infinity is represented only
by one point. In the complete de Sitter space the situation is quite
different and is discussed in the next section.

Although infrared corrections do not appear in the 1-point function
$\langle\varphi(t)^2\rangle$, they contribute to the two point
function $\langle\varphi(1) \varphi(2)\rangle$. To illustrate this
consider the limit when $\tau_1=\tau_2=\tau$ and
$x^2=(\vec{x}_1-\vec{x}_2)^2\gg\tau^2$. This corresponds to
$z\rightarrow - \infty$ (\ref{geodesic distance}). The bare Green's
function in this limit is given by
$$
G_0(z, \mu)\ =\ \frac{1}{\sqrt{-2 z}} \Big[ N(\mu) (-z)^{i\mu}\ +\
N(-\mu) (-z)^{-i\mu} \Big]
$$
The exact Green's function is equal to\footnote{To derive this
formula we can make a Fourier transform
$$
\int^{\frac{\mu}{\tau}} dq \cdot \tau \Big[\ A(\mu)A^\ast(-\mu)\
\Gamma\ (q\tau)^{2i\mu}\ +\ A(-\mu)A^\ast(\mu)\ \Gamma^\ast\
(q\tau)^{-2i\mu}\ +\ \big(|A(\mu)|^2+|A(-\mu)|^2\big)\ C\ \Big]\
e^{i q x}
$$ and retain only terms of the order $\lambda^2\log(-z)$ while
neglecting the terms of the order $\lambda^2$.}
\begin{equation}\nonumber
\begin{split}
G(z)\ =\ \Big[ 1\ +\ \frac{\lambda^2}{2}
\Big(B(\mu)-B(-\mu)\Big)\Big(|g(\mu)|^2\ -\ |g(-\mu)|^2
\Big)\log(-z) \Big]\ G_0(z,\mu+\delta\mu)\ =\\=\ \Big[1\ -\
\frac{\lambda^2}{4\mu}\big(1-e^{-2\pi\mu}\big)|g(-\mu)|^2\log(-z)\Big]
\ G_0(z,\mu+\delta\mu)
\end{split}
\end{equation}
We see that besides the infrared renormalization of mass, which we
ignore in the present paper, the bare Green's function is multiplied
by the function of $\log(-z)$. Thus, even in Poincare patch,
infrared corrections {\bf do appear} when the two points are
separated by a large geodesic distance. It would be interesting to
understand the consequences of this result for the inflationary
models in Poincare patch.

\section{Secular interactions and leading logarithms, complete $dS$ space}
In order to describe the global $dS$ space, we use the standard
metric $ds^2\!=dt^2-\cosh^2t(d\Omega_d)^2$. The eigenmodes for
the~Bunch-Davies vacuum are inherited from the sphere. To simplify
notations we write them for $d=1$:
$$
f_q(t)\propto\:P^{-q}_{-\frac{1}{2}+i\mu}(i\sinh t)
$$
where $q$ is an integer. These modes are selected by~the~condition
that they are regular when continued to~the~southern hemisphere
($t=-i\vartheta$; $\vartheta>0$).

The logarithmic divergences appear when $|q|\!\gg\!1$ and
$|t|\!\rightarrow\!\infty$. In this regions the Legendre functions
can be replaced by the~Bessel functions. We have:
$$
f_q(t)\xrightarrow[q\rightarrow\infty]{}\begin{cases}
\tau^{d/2}h(q\tau),\quad\tau=e^{-t},\quad t\rightarrow\infty;\\
\widetilde{\tau}^{d/2}h^*(q\widetilde{\tau}),\quad\widetilde{\tau}=e^{+t},\quad
t\rightarrow-\infty.\end{cases}
$$
As it should be, this is exactly the~doubled Poincare patch.

Let us use these modes to~calculate perturbative corrections
to~$\langle\varphi^2(n)\rangle$, assuming that the interaction
begins adiabatically in the far past, with
$\widetilde{\tau}\!=\!\varepsilon\!\rightarrow\! 0$, while
the~"observer" sits in the future at fixed~$\tau$. The~most
important contribution comes from the $+-$ term in the~Fig.5.
\begin{figure}[h]
\center{\includegraphics[width=110mm]{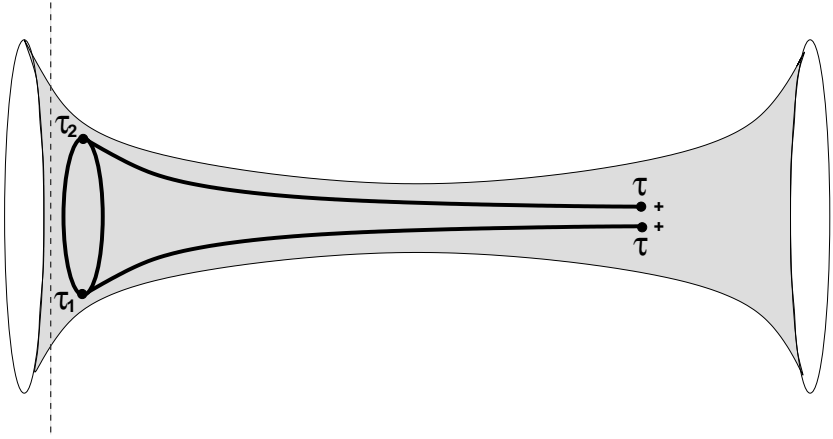}} \caption{{\footnotesize
Relevant diagram, leading to IR divergence, in complete $dS$
space.}}
\end{figure}
We have
\begin{equation}\label{one-point}
\langle\varphi^2(n)\rangle^{(1)}=\ \lambda^2 \tau^d\!\!\!\int
\frac{d^dq}{(2\pi)^d} |h(q\tau)|^2\int\limits_\varepsilon^\infty
\frac{d\widetilde{\tau}_1d\widetilde{\tau}_2}{\widetilde{\tau}_1\widetilde{\tau}_2}
\bigl(h^*(q\widetilde{\tau}_1)\,h(q\widetilde{\tau}_2)\bigr)
\!\cdot\!(\widetilde{\tau}_1\widetilde{\tau}_2)^{\frac{d}{2}}\!\cdot\!\sigma_q(\widetilde{\tau}_1,
\widetilde{\tau}_2)
\end{equation}
where
$$
\sigma_q(\widetilde{\tau}_1,
\widetilde{\tau}_2)=\int\frac{d^dk}{(2\pi)^d}\,h^*(k\widetilde{\tau}_1)\,h(k\widetilde{\tau}_2)
\!\cdot\!h^*(|k-q|\widetilde{\tau}_1)\,h(|k-q|\widetilde{\tau}_2).
$$
We consider here only the~dominant contribution, when $t_1$, $t_2$
are both in the far past. If $k\gg q$, we get the following scaling
property:
$$
\sigma_q(\widetilde{\tau}_1,
\widetilde{\tau}_2)\approx\sigma_0(\widetilde{\tau}_1,
\widetilde{\tau}_2)=(\widetilde{\tau}_1
\widetilde{\tau}_2)^{-d/2}\Phi\Big(\frac{\widetilde{\tau}_1}{\widetilde{\tau}_2}\Big).
$$
The integral (\ref{one-point}) becomes:
\begin{multline*}
\langle\varphi^2(n)\rangle^{(1)}=\ \lambda^2 \tau^d\!\!\!\int
\frac{d^d q}{(2\pi)^d}|h(q\tau)|^2\int\limits_\varepsilon^\infty
\frac{d\widetilde{\tau}_1d\widetilde{\tau}_2}{\widetilde{\tau}_1\widetilde{\tau}_2}
\bigg(B(\mu)\Big(\frac{\widetilde{\tau}_1}{\widetilde{\tau}_2}\Big)^{i\mu}+
B(-\mu)\Big(\frac{\widetilde{\tau}_1}{\widetilde{\tau}_2}\Big)^{-i\mu}
\bigg)
\Phi\Big(\frac{\widetilde{\tau}_1}{\widetilde{\tau}_2}\Big)=\\ =\
const\cdot \lambda^2 \cdot \tau^d\!\!\!\int d^dq |h(q\tau)|^2
\log\Bigl(\frac{\mu}{q\varepsilon}\Bigr).
\end{multline*}
The UV divergence at large $q$ must be cut-off by the condition
$q\tau\!\lesssim\!M_{pl}$. Thus we get the result
\begin{equation}\label{phi^2 complete space}
\langle\varphi^2(n)\rangle^{(1)}\ =\ const\cdot \lambda^2
M_{Pl}^{d-1}\log\Bigl(\frac{\mu}{M_{Pl}}\frac{\tau}{\varepsilon}\Bigr)\
=\ const\cdot \lambda^2 \langle\varphi^2(n)\rangle^{(0)}\cdot
\log\Bigl(\frac{\mu\tau}{M_{pl}\varepsilon}\Bigr).
\end{equation}
This formula is valid if:
$$
\varepsilon\ll\frac{\mu}{M_{Pl}} \tau
$$
which means that the time~$T$ during which the interaction was on,
satisfies
$$
T=\frac{1}{H} \log(\frac{\tau}{\varepsilon})\ \gg\
\frac{1}{H}\log\Big(\frac{M_{Pl}}{m}\Big)
$$
(where we reinstated the~Hubble constant).

In the Schwinger - Keldysh language we accounted for the $(+/-)$
self-energy part. There are, of course other insertions, $(+/+)$ and
$(-/-)$, also generating secular logarithms. However, they are
proportional to $\int d^{d}q\ h^{2}(q\tau )$ and its conjugate. This
integral is UV convergent due to the oscillations of $h(q\tau)$.
Hence there are no UV/IR mixing in these terms and their secular
contribution, while non-zero, does not contain $M_{pl}$, unlike
(\ref{phi^2 complete space}).

It is also instructive to write the above correction in the
covariant form. Various Schwinger-Keldysh propagators are expressed
in terms of the different boundary values of a single analytic
Wightman function, $g(n\cdot n^{\prime })$, e.g. $G_{++}=g(n\cdot
n^{\prime }-i0)$, $G_{+-}=g(n\cdot n^{\prime }+i\epsilon
sgn(n_{0}-n_{0}^{\prime }))$ etc. The function $g(z)$ is real for
$z\leq 1$ which corresponds to the space-like separations. By
combining terms in the Schwinger-Keldysh diagrams it is easy to get
$\langle \varphi ^{2}\rangle =\langle \varphi ^{2}\rangle
_{I}+\langle \varphi ^{2}\rangle _{II}$ where\footnote{The easiest
way to derive these formulas is to use the definition of
(anti)chronological products
$$
T(J_1 J_2)\ =\ \frac{1}{2}\{J_1,J_2\}\ +\ \sigma
\frac{1}{2}[J_1,J_2], \ \ \ \ \ \ \widetilde{T}(J_1 J_2)\ =\
\frac{1}{2}\{J_1,J_2\}\ -\ \sigma \frac{1}{2}[J_1,J_2],\ \ \ \ \ \
J_1 J_2\ =\ \frac{1}{2}\{J_1,J_2\}\ +\ \frac{1}{2}[J_1,J_2],
$$ where $\sigma=sign(n_{10}-n_{20})$ and then use the symmetry of the measure w.r.t. intrchange of $1\leftrightarrow
2$ to reduce the integration domain to the region $n_{10}>n_{20}$.
All the terms containing anticommutator of currents are collected
into $\langle\varphi^2\rangle_I$, all the terms containing
commutator are in $\langle\varphi^2\rangle_{II}$. }
\begin{equation}\nonumber
\begin{split}
\langle \varphi ^{2}\rangle_{I} \sim \int^{n_{0}} dn_{1} dn_{2}\
Im\Big(g(nn_{1})\Big)\ Im\Big( g(nn_{2})\Big)\ \langle \{J(n_{1}),J(n_{2})\}\rangle  \\
\langle \varphi ^{2}\rangle _{II} \sim -2i \int^{n_{0}}dn_{1}dn_{2}\
 \theta(n_{10}-n_{20})\ Im\Big(g(nn_{1})\Big)\ Re\Big(g(nn_{2})\Big)
\langle \lbrack J(n_{1}),J(n_{2})]\rangle
\end{split}
\end{equation}%
where $J\sim \varphi ^{2}(n).$ The first term represents the
contribution of the real particles created from the vacuum, while
the second term comes from the virtual particles. The logarithmic
divergence arises from the domain where $n_{10},n_{20}\rightarrow
-\infty$, while $(n_{1}n_{2})\sim 1$. The mathematical origin of the
UV/IR \ mixing lies in the fact that the first integrand contains
terms $g(z+i0)g(z^{\prime }-i0)$ which become singular on the light
cone while infrared divergent in the infinite past. This phenomenon
never happens in the Minkowski space.

In higher orders there are higher powers of the logarithms. Their
summation requires the kinetic equation and will be discussed
elsewhere.

\section{Conclusions}
The physical interpretation of the above estimates is the following.
We are considering a complete $dS$ space. All points of this space
are equivalent, so that the statements that at a given point we have
expansion or contraction are meaningless. However, if we fix the
position of the observer, one can define domains, such that the
signal sent from them will be either red shifted or blue shifted.
The essence of the formula (\ref{phi^2 complete space}) can be
grasped from the Fig.5. We integrate the interaction over the
faraway past region. The size of the loop determines the interaction
scale $\sim 1/m$, which is a large quantity. While the signal from
the interaction region propagates along the geodesics to the
observer, sitting at the point $\tau$, it is blue-shifted to the
Planck scale $\sim 1/M_{pl}$. As a result we get a very curious
UV/IR mixing. In the flat space we expect that UV and IR
divergencies contribute to the physical quantities independently -
we do not expect the terms, like (\ref{phi^2 complete space}), which
are both UV and IR divergent at the same time. This is a specific
feature of the curved space.

The $\varepsilon $ dependence of the physical matrix element
discussed above indicates a breakdown of the $dS$ symmetry; as
always, spontaneous symmetry breaking manifests itself through the
sensitive dependence on the boundary conditions. The logarithms will
be present even for a patch of the $dS$ space, provided that it is
"large", that is the past cone of the observer intersects a decent
portion of the past infinity. As we saw from Fig.4, this is not the
case for the Poincare patch; for it "the world is not enough".

Let us explain our motives for using the global $dS$ space, while in
the inflationary theories only a small part of it is usually
present. Our goal is to resolve the puzzle of the cosmological
constant by infrared means. We start with the Einstein action with
the cosmological constant present. The standard procedure in field
theory is to assume first that we can neglect quantum corrections at
large distances, find a classical solution and then evaluate the
corrections. It is this procedure which allows us to use classical
Einstein or Navier - Stokes and forbids the similar use of the Yang
- Mills equations (due to asymptotic freedom) and sometimes the
diffusion equation (due to Anderson's localization).

In such a setting we must consider the global $dS$ space as a first
step. If a starting point were incomplete space, we would end up
with the unitarity problem, since the particles can disappear from
the space. Of course it is possible to have a Poincare patch glued
to the Minkowski one in such a way that the result is geodesically
complete. However this space will not be a solution of the Einstein
equations with the cosmological constant only. It is also possible
to modify the Einstein action so that we have a different background
without IR divergences. This looks ambiguous and is far from our
goal, which is to tame infrared divergencies. We should remember
that to solve the $\Lambda$-problem one must be searching for the
infrared effects and not running from them. IR divergence is not a
problem but an opportunity.

Another question is related to the choice of the Bunch-Davies vacuum
in the above calculation. What is the reason for this (apart from
the tradition)? It seems that the right starting point should be the
state with the longest life time. We haven't proved that this is the
case, but various estimates make us believe that the Bunch-Davies
vacuum is the most stable one. In the appendix we present the
propagators for the different possible ground states. It should not
be difficult to extend our analysis to other vacua.

Finally, there is a number of valuable papers \cite{intr-dev}
intersecting with our work, but it seems that our approach brought
some new and unusual results.

\section{Acknowledgements}
We would like to thank E.Akhmedov, J.Maldacena and V.Mukhanov for
useful discussions. This work was supported in part by the NSF grant
number PHY-0756966.

\section*{Appendix A. Oscillator, relative probabilities}
Here we briefly discuss some ideas mentioned in the main text using
the simplest model - quantum mechanical particle. Take an oscillator
with variable frequency
$$
(\partial_t^2\ +\ m^2\ +\ U(t))\varphi\ =\ 0
$$
Let us introduce Jost functions
\begin{eqnarray}
f_{in}(t) \rightarrow \frac{1}{\sqrt{2 m}} e^{i m t} \ \ \ \ \ \
\ t\rightarrow -\infty \nonumber\\
f_{out}(t) \rightarrow \frac{1}{\sqrt{2 m}} e^{i m t} \ \ \ \ \ \ \
t\rightarrow +\infty
\end{eqnarray}
As well known in scattering theory
$$
f_{in}(t)= \alpha f_{out}(t)\ +\ \beta f_{out}^\ast(t)
$$
$|\alpha|^2-|\beta|^2=1$. Let us find the vacuum decaying amplitude.
We define the (in) and (out) vacua in a usual way:
$$
\varphi\ =\ a f_{in}^\ast + a^\dagger f_{in}\ =\ b f_{out}^\ast +
b^\dagger f_{out}
$$
and $a|0\rangle_{in}=b|0\rangle_{out}=0$, $[ a, a^\dagger] = [b,
b^\dagger]=1$.  The in/out Green's function
$$
G=_{\ out}\!\langle0| T \varphi(t_1)\varphi(t_2)|0\rangle_{in}
\frac{1}{_{out}\langle 0|0\rangle_{in}}\ =\ \frac{1}{\alpha}
f_{in}(t_<) f_{out}^\ast(t_>)
$$
This Green's function is the one satisfying the composition
principle \cite{Polyakov}
$$
\frac{\partial G(t_1,t_2)}{\partial m^2}\ =\ -i
\int\limits_{-\infty}^{\infty} G(t_1, t) G(t, t_2) dt
$$
This equation allows to represent $G$ in terms of the Feynman's sum
over paths. The amplitude to produce $2n$-particles is given by:
$$
A_{0\rightarrow 2n}\ =\ _{out}\langle 0| \frac{b^{2
n}}{\sqrt{2n!}}|0\rangle_{in}\ =\ _{out}\langle 2n|0\rangle_{in}
$$
Let us express it in terms of the Green functions:
$$
G(t_1,...,t_{2n})\ =\ \frac{_{out}\langle0|T
\varphi(t_1)...\varphi(t_{2n})|0\rangle_{in}}{_{out}\langle
0|0\rangle_{in}}\ \rightarrow_{t_j\rightarrow \infty}
\frac{_{out}\langle0|b^{2n}|0\rangle_{in}}{_{out}\langle0|0\rangle_{in}}
\frac{\Big(  e^{-i m \sum t_j} \Big)}{(2m)^n}+ ...
$$
The Wick theorem on another hand gives
$$
G(t_1,...t_{2n})\ =\ G(t_1,t_2)... G(t_{2n-1},t_{2n}) + perm.
$$
Total number of permutations is $(2n-1)(2n-3)...\ =\ (2n-1)!!\ =\
\frac{(2n)!}{n! 2^n}$. From here we derive
$$
\bigg|\frac{A_{0\rightarrow 2n}}{A_{0\rightarrow 0}}\bigg|\ =\
\frac{(2n)!}{2^n n!} \Big(\frac{\beta}{\alpha}\Big)^{n}
\frac{1}{\sqrt{(2n)!}}
$$
(we pick up the terms containing $e^{-im\sum t_j}$ from the products
of the Green's functions). Hence the probability to produce $2n$
particles is
$$
\frac{W_{2n}}{W_0}\ =\ \frac{\big|_{out}\langle
2n|0\rangle_{in}\big|^2}{|_{out}\langle0|0\rangle_{in}|^2}\ =\
\frac{(2n)!}{(n!)^2 4^n}\bigg|\frac{\beta}{\alpha}\bigg|^{2n}
$$
The normalization condition gives
$$
1\ =\ W_0\ +\ \sum\limits_{n=1}^\infty W_{2n} \ =\ W_0 \big(1+
\sum\limits_{n=1}^\infty\frac{W_{2n}}{W_0}\big)\ =\ W_0
\sum\limits_{n=0}^\infty \frac{(2n)!}{(n!)^2 4^n}
\bigg|\frac{\beta}{\alpha}\bigg|^{2n}\ =\
\frac{W_0}{\sqrt{1-\big|\frac{\beta}{\alpha}\big|^2}}
$$
Thus $W_0\ =\ \sqrt{1-|\frac{\beta}{\alpha}|^2}\ =\
\frac{1}{|\alpha|}$. It is interesting to note that the Green
functions give {\it relative } probabilities, $P_{0n}\ =\
\frac{W_{n}}{W0}$. When the vacuum is unstable $W_{0}<1$ and hence
$\sum\limits_{n}P_{0n}>1$. This may open the way to interpretation
of the non-unitary field theories - they describe unstable vacua.
For example in non-unitary CFT there are well classified operators
with negative norms. The 'probabilities' extracted from the Green
functions satisfy $\sum\limits_{n}\pm P_n=1$, $\sum\limits_{n}
P_n>1$ and $P_n$ can be interpreted as relative probabilities.

Another interesting relation is $W_{0}\ =\ e^{-\Gamma}$,
$\frac{\partial\Gamma}{\partial m^2}\ =\ \int Im\ G(t,t)\ dt\ =\ Im
\Big( \frac{\beta}{\alpha}\int f_{out}^\ast(t)^2 dt\Big)$. We see
that $Im\ G(t,t) \neq 0$ signals vacuum instability. The typical
back-reaction of produced particles on the field $U(t)$ is
characterized by the 'current'\ \  $J\ =\
_{in\!}\langle0|\varphi(t)^2|0\rangle_{in}$. We see that
$$
J(t)\ =\ |f_{in}(t)|^2
$$
and (omitting rapidly oscillating terms)
$$
J(\infty)-J(-\infty)\ =\ \frac{1}{2m} (|\alpha|^2+|\beta|^2-1)\ =\
\frac{1}{m} |\beta|^2
$$
Thus, the back-reaction does not become large with time. We can also
evaluate the average number  of the produced excitations
$\overline{n}\ =\ \sum W_{2n} (2n)\ = \ |\beta|^2$. However it is to
be remembered that this number fluctuates:
$\overline{(n-\overline{n})^2}\ \sim \ \overline{n}$.

\section*{Appendix B. The Green's functions of the de Sitter space.}
The propagator is a solution to the inhomogeneous Klein-Gordon
equation with a delta-function source. Such solution is not unique
and defined modulo a solution to homogeneous equation. In the flat
Minkowski space there is a well established prescription of doing
integration in the complex momentum plane which fixes this ambiguity
and uniquely defines the propagators. This prescription however is
not straightforwardly generalized to curved space which we are
working with, hence we need a different setup. A convenient method
is to write the propagator in the form
\begin{equation}\label{Wronskian formula}
G(t_1 , t_2)\ =\ \frac{1}{W[\varphi_1,\varphi_2]}
\varphi_1(t_<)\varphi_2(t_>)
\end{equation}
Where $\varphi_{1,2}(t)$ are two linearly independent solutions to
the homogeneous equation, $W[\varphi_1,\varphi_2]$ is their
Wronskian and $t_{<>}$ is the smallest and largest of the times
$t_{1,2}$. The above mentioned ambiguity is reflected here in the
different possible choices for the solutions to the homogeneous
equation $\varphi_{1,2}(t)$. This choice depends on the problem we
\begin{figure}[h]
\center{\includegraphics[width=145mm]{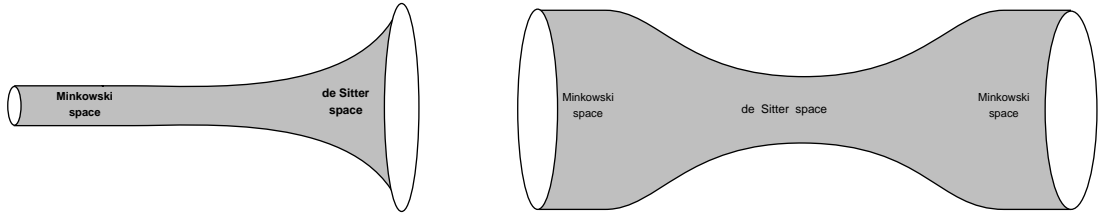}}
\caption{{\footnotesize On the left - geometry of expanding
Universe, on the right - geometry of complete de Sitter space.}}
\end{figure}
want to solve. In case of in-out propagator we should choose
$\varphi_{1,2}(t)$ to be Jost functions of the scattering problem,
i.e. $\varphi_1$ to be a plane wave at $t\rightarrow - \infty$ and
$\varphi_2$ to be a plane wave at $t\rightarrow + \infty$. In case
of in-in propagator both $\varphi_1$ and $\varphi_2$ should be plane
waves at $t\rightarrow - \infty$.

Another complication comes from the fact that a de Sitter patch can
be embedded into flat space in numerous ways. Below we illustrate
this by considering two possibilities: initially flat space
adiabatically starts to expand; and complete de Sitter space glued
with two cylinders on both plus and minus time infinities (see
Fig.6). The Green's functions for these geometries turn out to be
very different.

\subsection{First geometry, Poincare patch}
We start with the first geometry, which can be conveniently
described by the Poincare patch $ ds^2\ =
\Big(\frac{\alpha}{\tau}\Big)^2(d\tau^2\ -\ dx^2)$~and for
simplicity we work in $d=1$ dimensions. The wave equation for a mode
with comoving momentum $p$ can be written as
$$
\phi_{\tau\tau}\ +\ p^2 \phi\ +\ \frac{(m\alpha)^2}{\tau^2}\phi\ =\
0
$$
with $\nu\ =\ i\mu\ =\ i \sqrt{(m\alpha)^2\ -\ \frac{1}{4}}$. It's
solution can be written in terms of Bessel functions. In the
in-region the Jost functions are Hankel functions, while in the out
region they are Bessel functions. Thus, according to (\ref{Wronskian
formula}) we obtain
\begin{equation}\label{Bess-Hank decomposition}
G^{in/out}_{P}(1,2)\ =\ \int\limits_{0}^{\infty} dp\
\cos[p(x_2-x_1)]\ \sqrt{\tau_1\tau_2}\
H^{(1)}_{\nu}\big(p\tau_>\big) J_{\nu}\big(p\tau_<\big)
\end{equation}
for in-out (Feynman) propagator and
\begin{equation}\label{Hank-Hank decomposition}
G^{in/in}_{P}(1,2)\ =\ \int\limits_{0}^{\infty} dp\
\cos[p(x_2-x_1)]\ \sqrt{\tau_1\tau_2}\
H^{(1)}_{\nu}\big(p\tau_>\big) H^{(2)}_{\nu}\big(p\tau_<\big)
\end{equation}
for in-in propagator. Integrals w.r.t. momentum can be carried out
explicitly by using 6.672.4, 6.672.3 of \cite{Gradshteyn-Ryzhik}.
The result is
\begin{equation}\label{in-out-de Sitter-Q}
G^{in/out}_{P}(1,2)\ =\ \mathcal{Q}_{\nu-\frac{1}{2}}(z+i0)
\end{equation}
$$
G^{in/in}_{P}(1,2)\ =\
\frac{1}{\cos(\pi\nu)}\mathcal{P}_{\nu-\frac{1}{2}}(-z-i0)
$$
The first expression was suggested in \cite{Polyakov}, the second
one is the Bunch-Davies propagator. Both are expressed in terms of
geodesic distance
\begin{equation}\label{geodesic distance}
z=\frac{\tau_1^2+\tau_2^2-(x_2-x_1)^2}{2\tau_1\tau_2}
\end{equation}
To make the integrals (\ref{Bess-Hank
decomposition}),(\ref{Hank-Hank decomposition}) convergent we need
to shift $\tau_>$ into complex plane. This shift together with
$\frac{\partial z}{\partial\tau_>}>0$  determines $i0$ prescription
in the arguments of Legendre functions.

Another nice representation of in-out propagator can be obtained
from 6.669.3-4 of \cite{Gradshteyn-Ryzhik}
\begin{equation}\label{in-out-Schwinger-de Sitter}
G^{in/out}_{P}(1,2)\ =\ \int\limits_{0}^{\infty} dp\ \cos[px_{21}]
\sqrt{\tau_1\tau_2}\ e^{-i\pi\nu}\ \int\limits_{0}^{\infty}
\frac{ds}{\sinh s} e^{i p(\tau_1\ +\ \tau_2) \coth s}\
J_{2\nu}\bigg( \frac{2 p\sqrt{\tau_1\tau_2}}{\sinh s}\bigg)
\end{equation}
This expression are analogous to Schwinger proper time
representation of the Feynman propagator in electric field. It's
path integral derivation can be found in \cite{Free particle}. It is
also convenient for the calculation of the imaginary part of in-out
propagator at coincident points, which is related to the imaginary
part of effective action. Taking the limit of coincident points in
(\ref{in-out-Schwinger-de Sitter}) and integrating over $k$ by using
6.611.1 we get up to inessential constant
$$
Im \Big[G^{in/out}_{P}(1,1)\Big]\ =\ Im
\frac{-1}{2\pi}\int\limits_{-\infty}^{+\infty}\frac{ds}{\sinh s}
e^{2i \mu s}\ =\ - \sum\limits_{n=1}^{\infty}(-1)^n e^{-2\pi n\mu} \
=\ \frac{e^{-2\pi\mu}}{1+e^{-2\pi\mu}}
$$
Non-vanishing of this quantity signals instability of the vacuum
w.r.t. creation of particles, similarly to Schwinger mechanism in
constant electric field.

\subsection{Geometry of the Complete de Sitter space}
The second geometry can be described by a metric
$$
ds^2\ =\ dt^2\ -\ \cosh t^2 d\varphi^2
$$
with compact coordinate $\varphi\in[0,2\pi]$. The wave equation for
a mode with integer momentum $p$ can be written as
$$
\ddot{\phi}\ +\ \tanh t \dot{\phi}\ +\ m^2 \phi\ +\ \frac{p^2}{\cosh
t^2}\phi\ =\ 0
$$
The general solution of this equation is\footnote{In the following
we adopt the notation $\mathsf{P}_\nu^\mu(x)$ and
$\mathsf{Q}_\nu^\mu(x)$ for associated Legendre functions on the cut
when the argument is $-1<x<1$. These functions are defined by
\cite{Gradshteyn-Ryzhik} 8.702-8.705. For Legendre functions in the
complex plane we use symbols $\mathcal{P}_\nu^\mu(z)$ and
$\mathcal{Q}_\nu^\mu(z)$.}
$$
\phi\ =\ \frac{1}{\sqrt{\cosh t}} \Bigg[\ C_1
\mathsf{P}_{p-\frac{1}{2}}^{\pm\nu}(\pm\tanh t)\ +\ C_2
\mathsf{Q}_{p-\frac{1}{2}}^{\pm\nu}(\pm\tanh t)\ \Bigg]
$$
Choosing the Jost functions and plugging them into (\ref{Wronskian
formula}) we obtain
$$
G^{in / in}_{dS} \ =\ \frac{\pi}{2 \sqrt{\cosh t_1  \cosh t_2}}
\sum\limits_{p=0}^{\infty} \varepsilon_p \cos{(p \varphi)}
\mathsf{P}^{-\nu}_{p-\frac{1}{2}}(-\tanh t_>)
\mathsf{P}^{\nu}_{p-\frac{1}{2}}(-\tanh t_<)
$$
\begin{equation}
\begin{split}\nonumber
G^{in / out}_{dS}\ =\ \frac{1}{ \sqrt{\cosh t_1  \cosh t_2}}
\sum\limits_{p=0}^{\infty} \varepsilon_p \Gamma(\nu-p+\frac{1}{2})
\Gamma(\nu+p+\frac{1}{2}) \cos(p\varphi)
\mathsf{P}^{-\nu}_{p-\frac{1}{2}}(\tanh t_>)
\mathsf{P}^{-\nu}_{p-\frac{1}{2}}(-\tanh t_<)\ =\\ =\
\frac{2}{\cos(\pi\nu)\sqrt{\cosh t_1 \cosh t_2}}
\sum\limits_{p=0}^{\infty} \varepsilon_p \cos(p\varphi)
\mathsf{P}^{-\nu}_{p-\frac{1}{2}}(\tanh t_>)
\mathsf{Q}^{\nu}_{p-\frac{1}{2}}(\tanh t_<)
\end{split}
\end{equation}
where $\varepsilon_p=1$ for $p=0$ and $\varepsilon_p=2$ otherwise.
It is possible to sum up the $p$-series and express the result in
terms of Lorentz invariant quantities (geodesic distance $z$ and
$\sigma\ =\ Sign[n_0(1)\ +\  n_0(2)]$).
\begin{equation}\nonumber
\begin{split}
G_{dS}^{in/in}\ =\ \frac{1}{2 i}\Big[
\mathcal{Q}_{\nu-\frac{1}{2}}(-z-i\varepsilon)\ -\
\mathcal{Q}_{\nu-\frac{1}{2}}(-z+i\varepsilon)\Big]\ -\ \frac{\pi}{4
i
\cos(\pi\nu)}(\sigma+1)\Big[\mathcal{P}_{\nu-\frac{1}{2}}(z+i\varepsilon)\
-\ \mathcal{P}_{\nu-\frac{1}{2}}(z-i\varepsilon)\Big]
\end{split}
\end{equation}
We would like to emphasize that this expression is Lorentz
invariant, since functions $\mathcal{P}_{\nu-\frac{1}{2}}(z \pm
i\varepsilon)$ have a cut only for $z<-1$,  but $\sigma$ is Lorentz
invariant quantity for $z<-1$. Also, this function vanishes within
the light-cone of the past of the antipod\footnote{It could be
convenient to cover the manifold $n_0^2\ -\ n_1^2\ -\ n_2^2\ =\ 1$
by coordinates
$$ \Bigg\{\begin{array}c
n_0\ =\ \cot \theta \\
n_1\ =\ \frac{\cos \varphi}{\sin\theta} \\
n_2\ =\ \frac{\sin\varphi}{\sin\theta}
\end{array}\ \ \ \ \ \ \ \ \ \ z\ =\ \frac{\cos(\varphi_1-\varphi_2)\ -\ \cos\theta_1\cos\theta_2}{\sin\theta_1 \sin \theta_2}\ \ \ \ \ \ \ \ \ \ \ \ \sigma\ =\ Sign[\sin(\theta_1+\theta_2)]$$
}. This area corresponds to $z<-1$, $\sigma=-1$, see Fig.7.
\begin{figure}[h]
\center{\includegraphics[width=85mm]{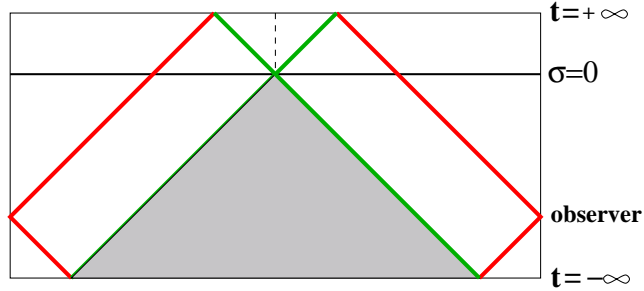}} \caption{{\footnotesize
Conformal diagram. Red - the light cone of the observer, green - the
light cone of the antipodal point. Gray - the area where
$G^{in/in}_{dS}=0$.}}
\end{figure}

For the Feynman propagator we obtain
$$
G^{in / out}_{dS}\ =\
 \frac{1}{\cos(\pi\nu)} \Big[
\mathcal{Q}_{\nu-\frac{1}{2}}(-z-i\varepsilon)\ +\
\mathcal{Q}_{\nu-\frac{1}{2}}(-z+i\varepsilon)\Big]
$$
Note, that this answer is different from in-out propagator in
Poincare patch (\ref{in-out-de Sitter-Q}), c.f. also
\cite{Emil-propagator}. It also has a non-vanishing imaginary part
at coincident points.

\end{document}